\documentclass[reprints,showpacs,superscriptaddress,amsmath,amssymb,prb]{revtex4} 

\usepackage[hypertex]{hyperref}
\usepackage{ifpdf}
\usepackage[english]{babel}
\usepackage{rotating}
\usepackage{caption}
\usepackage{graphicx}
\usepackage[dvips]{color}
\newcommand{\pfn}{Pb(Fe$_{1/2}$Nb$_{1/2}$)O$_{3}$\,}
\newcommand{\pfns}{PFN\,}
\newcommand{\musr}{$\rm \mu SR$\,}
\usepackage{graphicx}
\usepackage{dcolumn}
\usepackage{bm}
\begin{document}
\title
{Spin-glass state and long-range magnetic order in \pfn}
\author{
G.M. Rotaru
       }
\affiliation{
Laboratory for Neutron Scattering, ETH Zurich 
and Paul Scherrer Institut, CH-5232 Villigen, PSI
            }
\author{
B. Roessli
       }
\affiliation{
Laboratory for Neutron Scattering, ETH Zurich 
and Paul Scherrer Institut, CH-5232 Villigen, PSI
            }
\author{
A. Amato
       }
\affiliation{
Laboratory for Muon Spin Spectroscopy, Paul Scherrer Institut, CH-5232 Villigen, PSI
            }
\author{ 
S.N. Gvasaliya
       }
\affiliation{
Laboratory for Neutron Scattering, ETH Zurich 
and Paul Scherrer Institut, CH-5232 Villigen, PSI
            }
\author{
C. Mudry
       }
\affiliation{
Condensed matter theory group, 
Paul Scherrer Institut, CH-5232 Villigen, PSI
            }
\author{ 
S.G. Lushnikov
       }
\affiliation{
Ioffe Physical Technical Institute of Russian Academy of Science,
26 Politekhnicheskaya, 194021 St. Petersburg, Russia
            }
\author{
 T.A. Shaplygina
       }
\affiliation{
Ioffe Physical Technical Institute of Russian Academy of Science,
26 Politekhnicheskaya, 194021 St. Petersburg, Russia
            } 
%
%
\begin{abstract}
We have investigated the magnetic ground-state of the multiferroic relaxor ferroelectric \pfn with $\mu$SR 
spectroscopy and neutron scattering. We find that a transition to a partially disordered phase occurs 
below $T=20$~K that coexists with long-range antiferromagnetic ordering. The disordered phase resembles a 
spin-glass. No clustering of magnetic ions could be evidenced by $\mu$SR so that the coexistence 
appears homogeneous in the sample. 
\end{abstract}
\pacs{{75.50.Lk} {Spin glasses and other random magnets}; {77.80.-e} {Ferroelectricity and antiferroelectricity}; {61.05.F} {Neutron diffraction and scattering}; {76.75.+i} {Muon spin rotation and relaxation} 
}
\maketitle
\section{Introduction}

Since the discovery of a large magneto-electric effect in RMn$_2$O$_3$%
~\cite{kimura} and RMn$_2$O$_5$~\cite{hur} 
(R=Rare Earth), there is a revival of interest in the study of multiferroic 
materials where the coupling between the electric polarization and the 
magnetic order is strong. Although coexistence of ferroelectricity and 
magnetic long-range order is not very common, quite a number of new 
multiferroics materials have been found recently, like Ni$_3$V$_2$O$_8$,%
~\cite{lawes}
MnWO$_4$,%
~\cite{taniguchi} 
or LiCuVO$_4$,%
~\cite{schrettle} 
to cite only a few. Multiferroics that have coupled ferroelectric and magnetic
order parameters are also promising materials for applications, 
as it is possible to control the ferroelectric polarization 
by a magnetic field and  vice-versa.~\cite{ramesh} 
Therefore, there is an extensive search for materials 
that show pronounced magneto-electric effect at ambient conditions. 
B-site disordered complex perovskites PbB$'_{x}$B$''_{1-x}$O$_3$ might be 
attractive candidates as they can adopt various ions which strongly influences 
the temperature of magnetic and ferroelectric ordering. 

Lead iron niobate, \pfn (\pfns), is one of the well known complex perovskites 
that exhibits multiferroic properties.~\cite{schmid} 
\pfns is rhombohedral at room temperature and cubic above 
the ferroelectric Curie temperature of 383~K.~\cite{yokomizo} 
\pfn undergoes a diffuse phase transition without pronounced frequency 
dispersion.~\cite{radhika} 
In the cubic phase, the symmetry of the chemical cell is $Pm\bar 3 m$. 
There is a cubic-tetragonal phase transition at $T=383$~K 
followed by a first-order transition at 354~K. 
The symmetry of the chemical structure of the low-temperature 
phase of \pfns was the subject of some debate. 
Lampis {\it et al.}~\cite{lampis} reported that the 
structure is monoclinic with space group $Cm$, 
whereas Ivanov {\it et al.}~\cite{ivanov} using powder neutron 
diffraction data found that the chemical structure of \pfn has 
$R3c$ symmetry at $T=10$~K.

The magnetic properties of \pfns are also not completely understood. 
\pfns undergoes a paramagnetic to antiferromagnetic 
phase transition  at $T_N\sim$ 140-160~K.~\cite{bokov} 
The magnetic structure was determined by single crystal neutron 
diffraction and corresponds to a simple $G$-type antiferromagnetic 
arrangement of the magnetic moments at all 
temperature below $T_N$.~\cite{howes} 
Magneto-electric effect in \pfns at low temperatures has been demonstrated 
long ago~\cite{watanabe} and, 
recently, it was shown that the DC magnetic susceptibility has an anomaly at 
the paraelectric-ferroelectric phase transition.~\cite{blinc20071} 
The dielectric permeability also shows an 
anomaly at the N\'eel temperature~\cite{yang}, 
which suggests that a biquadratic term in the Landau free energy of 
the form $P^2M^2$ is responsible for the coupling between the ferroelectric 
polarization $P$ and the (staggered) magnetization $M$. 
Additional magnetic anomalies were reported in the literature for \pfn below 
the magnetic phase transition. Both NMR ~\cite{blinc2008} 
and susceptibility measurements~\cite{bhat} 
have shown that an additional magnetic phase transition occurs in \pfns  
around $T_g\sim$20~K that was interpreted as the formation of 
a spin-glass phase by Kumar {et al.}~\cite{kumar2008}
 
Here we report on muon-spin rotation and neutron scattering results 
of the magnetic properties of PFN below the N\'eel temperature. 
We show that the magnetic ground-state of \pfn is a spin-glass-like  
state that coexists with long-range antiferromagnetic order below 
$T_g\simeq20$~K. 
 
\section{Experimental}

The \pfn single crystals used in the muon spin rotation 
and neutron scattering experiments were grown by spontaneous 
crystallization from the melt following the routine described in 
Ref.~\onlinecite{kumar2008}. 
The muon spin relaxation measurements were performed using the GPS 
instrument at the Paul Scherrer Institut (Villigen, Switzerland) 
on a \pfn single crystal between 160~K and 5~K. 
The data were recorded using the zero-field method that allows to 
determine both the static and dynamics in disordered spins systems. 
The neutron scattering experiments were performed with the cold neutron 
three-axis spectrometer TASP~\cite{tasp} at SINQ operated in diffraction mode 
at $k_f$=1.97~\AA $^{-1}$. For this experiment a single crystal of \pfn 
was oriented with the [1,1,0] and [0,0,1] crystallographic directions 
in the scattering plane. In that scattering geometry, magnetic Bragg peaks 
with indexes $(h/2,h/2,l/2)$  in the rhombohedral setting can be accessed. 
The use of a three-axis spectrometer was justified by the need to improve 
the ratio intensity/background in order to 
search for possible short-range order at low temperatures. 

\section{Results}

\subsection{\musr}

In a \musr experiment, the time evolution of the muon spin polarization 
is monitored by recording the asymmetric emission 
of positrons produced by the weak decay of the muon (muon lifetime 
$\sim 2.2~\mu$s). The time histogram of the collected positrons 
is given by
\begin{equation}
N(t)=N_0\exp(-t/\tau)[1+AG_z(t)] + B
\end{equation} 
where $A$ is the initial muon asymmetry parameter, 
$N_0$ is a normalization constant, and $B$ is a time-independent background. 
The function $G_z(t)$ reflects the normalized muon-spin auto-correlation 
function 
\begin{equation}
G_z(t)=
\frac{\langle\vec S(t)\cdot\vec S(0)\rangle}
     {{\vec S}^2(0)}
\end{equation}  
where $\vec S$ is the spin of the muon. Hence $AG_z(t)$, 
often called the $\mu$SR signal, 
reflects the time evolution of the muon polarization.

\begin{figure}[h]
\begin{center}
\includegraphics[width=0.4\textwidth, angle=0]{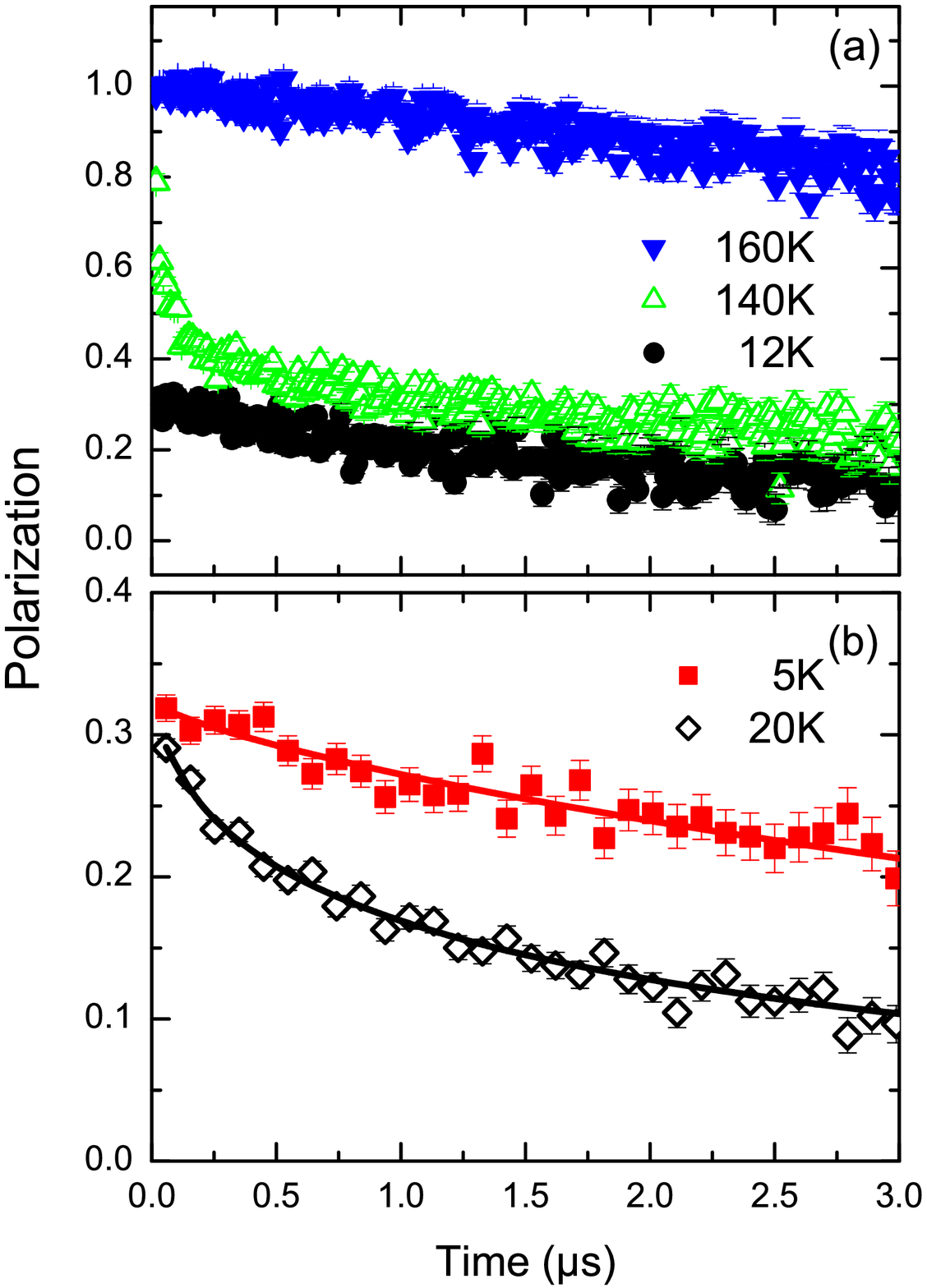}
\caption{
(Color online:)
\label{fig1}
(a) Time evolution of the muon-spin polarization reported for 
characteristic temperatures. Note the weak depolarization 
in the paramagnetic phase (160~K) and the loss of the signal amplitude 
below $T_N$, which retains the value 1/3 ($t = 0$) at low temperatures. 
(b) Examples of polarization signals recorded close to (20 K)
and well below (5 K) $T_g$. 
The lines represent fits performed with Eq.~(\ref{eq3}) 
(data reported on (b) are strongly binned for clarity).
        }
\end{center}
\end{figure}

{Figure}~\ref{fig1} shows the $\mu$SR signal in \pfn 
upon passing the N\'eel temperature.
In the paramagnetic phase the signal is consistent with 
a weak depolarization of the muon signal solely due to nuclear dipole moments. 
Below the N\'eel temperature, 
the muon signal changes rapidly and the initial asymmetry $A$ 
is strongly reduced. This indicates that 
there is a large distribution 
of internal fields seen by the $\mu^+$, 
which is a consequence of the random distribution of the Fe and Nb ions in the 
chemical lattice. Therefore,  
the $\rm \mu SR$ signal is depolarised and the residual asymmetry is 
close to 1/3 of the value observed in the paramagnetic 
state. Hence, in the long-range antiferromagnetic phase,  
the muon signal reflects 
only muons having their initial polarization 
along the direction of the internal field at the stopping site.

\begin{figure}[h]
\begin{center}
\includegraphics[width=0.4\textwidth, angle=0]{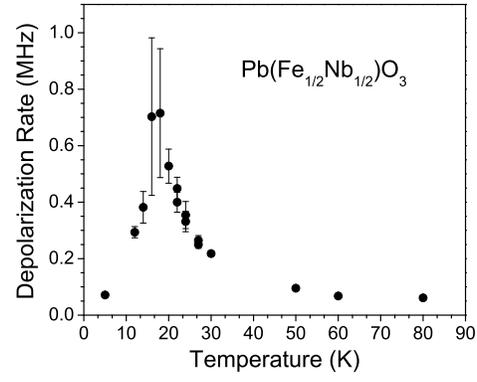}
\caption{
\label{fig2}
Temperature dependence of the $\mu$SR depolarization rate 
$\lambda$ below 80~K.
        }
\end{center}
\end{figure}

Below $T_N$, the time evolution of the remaining $\mu$SR signal is 
best fitted assuming the function
\begin{equation}
G_z(t)=\frac{1}{3}\exp[-(\lambda t)^\beta].
\label{eq3}
\end{equation}
Close to $T_N$, the power exponent is $\beta \simeq 1$, 
reflecting the presence of fluctuating internal magnetic fields with 
a unique correlation time $\tau_c$. In this range of temperatures, 
the depolarization rate $\lambda$ is proportional to 
the second moment of the magnetic field distribution 
$\langle B^2 \rangle$ and to the correlation time $\tau_c$ of 
the internal field and is given by 
\begin{equation}
\lambda=\gamma^2_\mu\langle B^2\rangle\tau_c~,
\label{eq4}
\end{equation}
\noindent where $\gamma_\mu$ is the muon gyromagnetic ratio. 
The temperature dependence of the depolarization rate $\lambda$ is shown 
in Fig.~\ref{fig2}. 
It can be observed that there is a critical-like divergence close to 
$T_g=20$~K. 
This is is a clear evidence that \pfn undergoes a second magnetic phase 
transition at the temperature $T_g$.

\begin{figure}[h]
\begin{center}
\includegraphics[width=0.4\textwidth, angle=0]{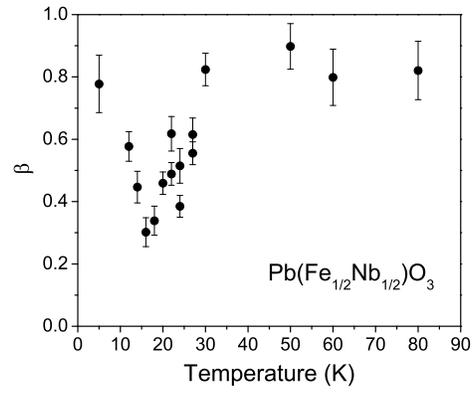}
\caption{
\label{fig3}
Temperature dependence of the parameter $\beta$, 
showing the occurrence of a 
distribution of relaxation times in \pfns (see text).
        }
\end{center}
\end{figure}

Figure~\ref{fig3} depicts the temperature evolution of the $\beta$ exponent. 
As already mentioned, at temperatures just below $T_N$, the muon 
depolarization is described by an exponential function
instead of a stretched exponential, 
i.e., with the parameter $\beta\sim$1. 
Upon further cooling below $T_N$,
$\beta$ continuously decreases and approaches the value of 1/3 
at $T_g\sim 20 $~K. A decrease of the exponent $\beta$ toward the value of 1/3 
has been predicted theoretically and later observed for the dense 
spin-glass systems AgMn and AuFe alloys.~\cite{campbell}
It is tempting to attribute this change of $\beta$ 
to the change in the distribution of the correlation times
$\tau_c$ of the internal magnetic fields at the muon stopping sites
as a spin-glass transition is approached from above. However, 
for an usual paramagnetic to spin-glass transition, 
one observes that the residual fluctuations lead to a constant value of 
$\beta$ below $T_g$ while $\lambda$ decreases, reflecting the decrease of 
the magnitude of the fluctuating field, as the temperature is lowered deeper 
into the spin-glass phase. However, Fig.~\ref{fig3} also shows that $\beta$ 
grows as the temperature is lowered below $T_g$ to recover the value 
$\sim 1$ at $T=2$~K. The solution of this apparent paradox is that the 
slow spin fluctuations, arising from the spin-glass state and which persist 
in a relatively broad temperature range below $T_g$, 
occur on top of antiferromagnetic fluctuations due to the long-range order. 
It is then natural to expect that the strong reduction of
the critical spin-glass fluctuations in time when the temperature
is decreased below $T_g$ leads to the recovery of the weak non-critical 
antiferromagnetic 
fluctuations probed by the $\rm \mu SR$ signal above $T_g$.
We note that as the static component of the $\mu$SR signal is already 
lost above $T_g$, no indication of a change (i.e., increase) of the damping 
of this component (as evidenced for example in other reentrant spin-glasses%
~\cite{mirebeau}) could be determined at the spin-glass transition by $\mu$SR. 

\subsection{Neutron scattering}

\begin{figure}[h]
\begin{center}
\includegraphics[width=0.4\textwidth, angle=0]{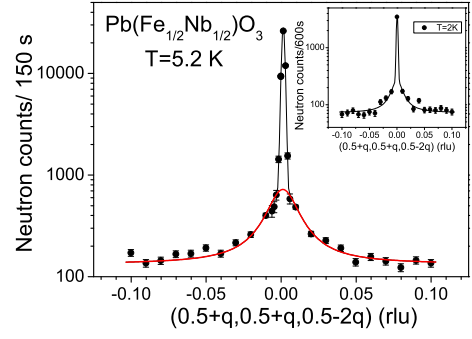}
\caption{
\label{fig4}
(Color online:)
Transverse scans through the 
${\bf Q}_{N}\equiv(0.5, 0.5, 0.5)$ 
magnetic Bragg peak. The broad component in 
the spectrum was fitted by a Lorentzian and is emphasized by a bold line. 
The Bragg peak is a fit to a Gaussian. 
The inset shows the intensity recorded in the neutron spin-flip channel 
for incident polarisation along the scattering vector.
        }
\end{center}
\end{figure}

Figure~\ref{fig4} 
shows a typical neutron elastic scan along the 
$(q,q,-2q)$ direction at $T=5.2$~K in \pfn. The 
spectrum consists of a sharp Bragg peak centered around 
${\bf Q}_{N}\equiv(1/2,1/2,1/2)$
that corresponds to the antiferromagnetic long-range order 
detected by neutron diffraction previously~\cite{howes}. 
A diffuse component is also present in the neutron spectrum that 
is weaker in intensity and has a width in $q$ that is much broader 
than the experimental resolution. 
The magnetic origin of the diffuse component was confirmed 
using the neutron polarization analysis 
device MuPAD~\cite{janoschek} that allows a complete separation 
of magnetic and nuclear scattering. Namely, using 
a neutron polarized beam for which the polarization is 
aligned along the scattering vector, magnetic scattering reverses 
the neutron polarization vector, 
whereas scattering of nuclear origin leaves the polarization 
of the neutron beam unchanged. 
Magnetic scattering therefore appears in the spin-flip channel, 
as shown to be the case for the diffuse component in the 
inset of Fig.~\ref{fig4}.

\begin{figure}[h]
\begin{center}
\includegraphics[width=0.4\textwidth, angle=0]{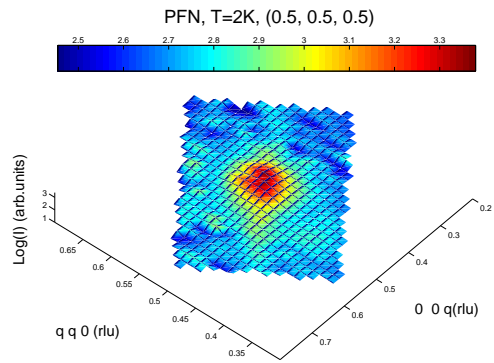}
\caption{
\label{fig5}
(Color online:)
Distribution of the elastic diffuse scattering measured in \pfn at 
$T=2$~K. For clarity the intense 
Bragg peak at ${\bf Q}_{N}\equiv(0.5, 0.5, 0.5)$  
was removed and the background subtracted from the data.
        }
\end{center}
\end{figure}

We investigated the distribution of the diffuse scattering around the  
${\bf Q}_{N}$ antiferromagnetic Bragg peak and found 
that it is essentially isotropically 
distributed in the (1,1,0)/(0,0,1) 
scattering plane accessible in the present measurements 
(see Fig.~\ref{fig5}).   
We performed an energy scan through the diffuse component and found 
that it is resolution-limited within the energy window 
($\Delta E$=0.3~meV) of the spectrometer. Therefore, 
the spin fluctuations associated with the diffuse scattering 
are static in the the time-window of our experiment.

\begin{figure}[h]
\begin{center}
\includegraphics[width=0.4\textwidth, angle=0]{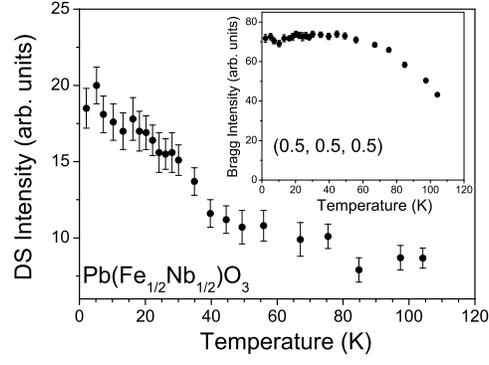}
\caption{
\label{fig6}
Temperature dependence of the integrated intensity 
of the diffuse scattering in \pfns 
measured around ${\bf Q}_{N}\equiv(0.5, 0.5, 0.5)$ Bragg peak. 
The temperature dependence of the 
antiferromagnetic Bragg peak is shown in the inset.
        }
\end{center}
\end{figure}

Within the quasi-static approximation,
 the energy-integrated neutron cross section is proportional to 
the generalized static susceptibility $\rm \chi({\bf Q})$~\cite{degenes} 
\begin{equation}
\begin{split}
I({\bf Q}, T)\propto&\, 
T\sum_{\alpha\beta}
(\delta_{\alpha\beta}-{\bf \hat Q}_{\alpha}{\bf\hat Q}_{\beta})
\sum_{dd'}F_d({\bf Q})F_{d'}({\bf Q})\chi^{dd'}_{\alpha\beta}({\bf Q})
\\
&\,\times
\exp\Big(i{\bf Q}({\bf r}_d-{\bf r}_d')\Big).
\end{split}
\end{equation}
The term $(\delta_{\alpha\beta}-{\bf \hat Q}_{\alpha}{\bf\hat Q}_{\beta})$ 
accounts for the fact that only fluctuations perpendicular 
to the scattering vector $\bf Q$ contribute to the diffuse intensity.
The index $d$ labels the spins at position ${\bf r}_{d}$ and $F_{d}({\bf Q})$ is the magnetic form factor.
We modeled the diffuse scattering measured along 
$(q,q,-2q)$ around the ${\bf Q}_{N}$ 
magnetic Bragg peak by a Lorentzian function
\begin{equation}
I({\bf Q},T)=
\frac{A(T)}
     {\kappa^2+|{\bf Q}-{\bf Q}_{N}|^2}
\label{eq5}
\end{equation}
\noindent where $A(T)$ is the scattered intensity and $\kappa=1/\xi$ is 
the inverse of the correlation length. 
The temperature dependence of the diffuse scattering intensity 
is shown in Fig.~\ref{fig6}. 
Above $T\sim$120~K the diffuse scattering is too weak to be separated 
from the background scattering. 
Below that temperature $A(T)$ increases monotonically 
with cooling down to $T\sim$2~K. 
A fit to the neutron data with Eq.~(\ref{eq5}) 
yielded at all temperatures, for which the diffuse scattering could be observed, 
$\kappa\sim 0.015 $~(rlu) that corresponds to $\xi\sim 17$~\AA. 

\section{Conclusion and discussion}

Both $\mu$SR spectroscopy and neutron diffraction have shown that 
a transition to a spin-glass state that coexists with long-range 
antiferromagnetic order occurs in the relaxor perovskite \pfn  
at $T_g\simeq 20$~K. Figure~\ref{fig1} 
shows that the asymmetry of the muon signal is rapidly lost below 
$T_N$ and does not recover when passing the spin-glass 
transition. This indicates that the magnetic-order is homogeneous and that 
it persists upon cooling the sample below $T_g$. 
This is confirmed by the presence of the resolution-limited 
magnetic Bragg peak at the lowest temperature showing that long-range 
antiferromagnetic order of the Fe$^{3+}$ spins is still present below 
$T_g$. The long-range antiferromagnetic order coexists with the disordered spin-glass-like phase. 
The signature of the spin-glass order appears in the neutron spectrum in the form of diffuse scattering.  

There are numerous examples of coexistence between the
ferromagnetic and spin-glass states of matter~\cite{fischer}.
For example, ferromagnetic and spin-glass orders 
are known to coexist in AuFe alloys at sufficiently low temperatures~\cite{murani}. 
Antiferromagnetic and spin-glass orders are also known to coexist
at sufficiently low temperatures, e.g.,
in Fe$_{0.55}$Mg$_{0.45}$Cl$_2$~\cite{shapiro1985}. 
The coexistence of collinear magnetic long-range order,
e.g., ferromagnetic or antiferromagnetic,
and spin-glass order is well understood theoretically 
at the level of the Sherrington-Kirkpatrick (SK)
\cite{Sherrington}
infinite-range model for classical 
(Ising- or Heisenberg) spins.
In the SK model, this coexistence is the rule
below some critical temperature and down to the lowest temperatures 
as soon as the magnitude $|\bar{J}|$ 
of the mean value of the random exchange coupling 
$J$ between any two spins is of the order of or larger than 
the standard deviation $\Delta{J}$  of its Gaussian distribution.%
~\cite{gabay,Korenblit,fyodorov,Liarte}

Classical spin models with short-range but random two-body spin interactions
(the Edwards-Anderson (EA) model of spin glasses~\cite{Edwards})
are not as well understood as the SK model. Monte Carlo simulations
suggest that the lower critical dimension for the 
onset of spin-glass order
when the disorder is dominant,
$|\bar{J}|/\Delta{J}\ll1$,
is strictly below three dimensions for Ising spins,~\cite{ballesteros}
a prerequisite for the coexistence of collinear and spin-glass
 order when $|\bar{J}|/\Delta{J}\gtrsim 1$.\cite{beath}
Fluctuation effects are stronger for Heisenberg spins 
than they are for Ising spins,
as fluctuations in the directions transverse to the 
direction taken by the collinear order parameter in spin space
become available in the former case.
Correspondingly, state of the art Monte Carlo simulations
give ``only'' the upper bound of three dimensions for
the spin-glass lower critical dimension
when the disorder is dominant,
$|\bar{J}|/\Delta{J}\ll1$.\cite{young} 
In other words, it cannot be ruled out, 
within the accuracy of available Monte Carlo simulations,
that the spin-glass phase is quasi-long-range ordered 
instead of long-range ordered 
when the classical spins are Heisenberg-like
in the three-dimensional EA model.

Given the uncertainty of
the spin-glass lower-critical dimension
for Heisenberg EA spin-glass models,
the mere observation that antiferromagnetic and spin-glass orders
coexists in \pfn is not sufficient to establish whether the relevant
magnetic degrees of freedom are Ising- or Heisenberg-like. 
Although exchange interactions remain to be determined in \pfn, 
the relatively high N\'eel temperature of \pfn suggests 
that exchange interactions dominate dipolar 
and single-ion anisotropies in this compound.
If this is the case, then \pfn 
is a good candidate to study the coexistence of long-range 
antiferromagnetic and spin-glass orders
for the Heisenberg spins of a three-dimensional 
antiferromagnetic EA model.

\section{Acknowledgements}

This work was performed at the spallation neutron source SINQ and the muon source S$\mu$S,
Paul Scherrer Institut, Villigen (Switzerland) and was partially
supported by the Swiss National Foundation (Project
No. 20002-111545). 

\section*{} 

\newpage
%
\newpage
%
%
%
\newpage
%
\newpage
%
\newpage
\newpage
%
\end{document}